\documentclass{appolb}
\usepackage{graphicx}

\begin{document}
\title{Scalar photoproduction on the proton at CLAS and GlueX energies%
\thanks{Presented at Excited QCD 2014}%
}
\author{M. L. L. da Silva
\address{Instituto de F\'{\i}sica e Matem\'atica, Universidade Federal de
Pelotas, Caixa Postal 354, CEP 96010-090, Pelotas, RS, Brazil.}
\\
{M. V. T. Machado
}
\address{Instituto de F\'{\i}sica, Universidade Federal do
Rio Grande do Sul, Caixa Postal 15051, CEP 91501-970, Porto Alegre, RS, Brazil.}
}
\maketitle
\begin{abstract}

In this work we present the results of a theoretical analysis of the data on
photoproduction of $f_0(980)$ meson in the laboratory photon energy between $3.0$ GeV and $3.8$ GeV.
A comparison is done to the measurements performed by the CLAS collaboration at JLab
accelerator for the exclusive reaction $\gamma p \to p f_0(980)$. The differential and integrated total
cross sections are also computed for the cases of the mesons $f_0(1500)$, and $f_0(1710)$, focusing on
the GlueX energy regime with photon energy $E_\gamma = 9$ GeV. 

\end{abstract}

\section{Introduction}

The spectroscopy of the low mass scalar mesons is an exciting issue in hadronic physics and is still an
unsettled question. Such a conflicting interpretation comes from the fact the situation is complex in low
energies where quantitative predictions from QCD are challenging and rely mostly on numerical techniques
of lattice QCD. In this context, the photoproduction of exotic mesons \cite{DK} can be addressed using
arguments based on vector meson dominance where the photon can behave like an $S=1$ quark-antiquark system.
Therefore, such a system is more likely to couple to exotic quantum number. Such a process could
provide an alternative to the direct observation of the radiative decays at low energies. Along those
lines, the GlueX experiment \cite{GlueX} is being installed and its primary purpose is to understand the
nature of confinement in QCD by mapping the spectrum of exotic mesons generated by the excitation of the
gluonic field binding the quarks. The mesons $f_0(1500)$ and $f_0(1710)$ are considered good candidates for
the scalar glueball \cite{Lee,Kirk}. However, in this mass region, the glueball state will mix strongly with
nearby $q\bar{q}$ states \cite{close1,close2}. The unknown about the mixing parameters still remains, on this
way some proposal to set the parameters is very important to determine the structure of this resonances.

The scenarios discussed in this paper consider the $f_0(980)$ as a tetraquark or as a ground state nonet.
We focus on the S-wave analysis on the forward photoproduction of $\pi^+\pi^-$ on the final state. We also
investigate the mixing in $f_0(1500)$ and $f_0(1710)$ resonances. The theoretical formalism employed is the
Regge approach with reggeized $\rho$ and $\omega$ exchange \cite{SM}. This assumption follows from Regge
phenomenology where high-energy amplitudes are driven by $t$-channel meson exchange.

\section{Model and cross section calculation}

According to the Regge phenomenology, one expects that only the $t$-channel meson exchanges are important in such a case. 
The reaction proposed here is $\gamma p \rightarrow p\,f_0(M)$. In the context of the Regge phenomenology the the narrow-width 
differential cross section for a meson of mass $m_S$ is given by \cite{SM},
\begin{eqnarray}
\frac{d\hat{\sigma}}{dt}(\gamma p \rightarrow p M)= \frac{|{\cal M}(s,t)|^2}{64\pi\,(s-m_p^2)^2},
\label{dsigma}
\end{eqnarray}
where ${\cal M}$ is the scattering amplitude for the process, $s,\,t$ are usual Mandelstan variables and $m_p$ is the proton mass.
For the exchange of a single vector meson, i.e. $\rho$ or $\omega$ one has:
\begin{eqnarray*}
&|{\cal M} (s,t)|^2 = - \frac{1}{2}{\cal A}^2(s,t)\left[\frac{}{}s(t-t_1)(t-t_2) 
+\frac{1}{2}t(t^2 - 2 (m_S^2 +s)t + m_S^4) \right] \nonumber\\
& - {\cal A}(s,t){\cal B}(s,t)m_ps(t-t_1)(t-t_2) 
-\frac{1}{8}{\cal B}^2(s,t)s(4m_p^2-t)(t-t_1)(t-t_2).
\label{msquare}
\end{eqnarray*}
where $t_1$ and $t_2$ are the kinematical boundaries
\begin{eqnarray}
\!\!\!\!\!\!t_{1,2} &=& \frac{1}{2s}\left[-(m_p^2-s)^2+m_S^2(m_p^2+s)\right. \nonumber\\
& \pm & \left. (m_p^2-s)\sqrt{(m_p^2-s)^2-2m_s^2(m_p^2+s)+m_S^4}\, \right],
\label{t12}
\end{eqnarray}
and where one uses the standard prescription for Reggeising the Feynman propagators  assuming a linear Regge trajectory
$\alpha_V(t)= \alpha_{V0}+ \alpha^\prime_V t$ for writing down the quantities ${\cal A}(s,t)$ and ${\cal B}(s,t)$:
\begin{eqnarray}
{\cal A}(s,t) & = & g_A\,\left(\frac{s}{s_0}\right)^{\alpha_V(t)-1}\frac{\pi\alpha^\prime_V}{\sin(\pi\alpha_V(t))}
\frac{1-e^{-i\pi\alpha_V(t)}}{2\,\Gamma(\alpha_V(t))},\nonumber \\
{\cal B}(s,t) & = & -\frac{g_B}{g_A}\,{\cal A}(s,t).
\label{abdef}
\end{eqnarray}
It is assumed non-degenerate $\rho$ and $\omega$ trajectories $\alpha_{V}(t) = \alpha_{V}(0)+\alpha^{\prime}_{V}t$, with
$\alpha_{V}(0)=0.55\,(0.44)$ and $\alpha^{\prime}_{V} = 0.8\,(0.9)$ for $\rho$ ($\omega$). In Eq. (\ref{abdef}) above, one has that
$g_A = g_S(g_V+2 m_p g_T)$ and $g_B=2g_Sg_T$. The quantities $g_V$ and $g_T$ are the $VNN$ vector and tensor couplings, $g_S$ is the
$\gamma V N$ coupling. For the  $\omega N N$ couplings  we have set $g_V^{\omega} = 15$ and $g_T^\omega =0$ \cite{SM} and for the
$\rho N N$ couplings we used $g_V^{\rho} = 3.4$, $g_T^{\rho}= 11$ GeV$^{-1}$. The $S V \gamma$ coupling, $g_S$, can be obtained from
the radiative decay width through \cite{KKNHH}
\begin{eqnarray}
\Gamma(S \to \gamma V) = g_S^2\frac{m_S^3}{32\pi}\Bigg(1-\frac{m_V^2}{m_S^2}
\Bigg)^3.
\label{width}
\end{eqnarray}
This model was applied on $f_0(1370)$, $f_0(1500)$ and $f_0(1710)$ mesons which was considered as mixing of $n\bar{n}$, $s\bar{s}$
and glueball states \cite{CDK}. In this case their radiative decays to a vector meson are expected to be highly sensitive to the
degree of mixing between the $q\bar{q}$ basis and the glueball. 
The numerical values for the widths having effects of mixing on the radiative decays of the scalars on $\rho$ and $\omega$
can be found in Table 1 of Ref. \cite{SM}. On this way it is clear that the
width is strongly model dependent and different approaches must be taken into account. For instance, we quote the work in Ref. 
\cite{Jacosa}, where the decays of a light scalar meson into a vector mesons and a photon, $S\rightarrow V\gamma$, are evaluated in
the tetraquark and quarkonium assignments of the scalar states. The coupling now reads,
\begin{eqnarray}
\Gamma(S \to \gamma V) = g_S^2\frac{(m_S^2 - m_V^2)^3}{8\,\pi\, m_S^3}.
\label{width_g}
\end{eqnarray}
The different nature of the couplings corresponds to distinct large-$N_c$ dominant interaction Lagrangians. For more details on this
calculation see Refs. \cite{SM,mm}.

\section{Results and discussions}

In what follows we present the numerical results for the direct $f_0(980)$ photoproduction considered in present study and the
consequence of the tetraquark and quarkonium assignments of the scalar states discussed in previous section and in Ref. \cite{mm}.
The results presented here will consider five distinct scenarios, three of them assuming that $f_0(980)$ is a quarkonium and two
assuming that $f_0(980)$ is a tetraquark. In scenarios 1, 2 and 3 the $f_0(980)$ will be interpreted as a ground-state nonet and in
scenarios 4 and 5 as a tetraquark. The $g_S$ coupling can be obtained from the radiative decay width in Table 1 of Ref. \cite{mm}
using Eq. (\ref{width}) for scenario 1 and using Eq. (\ref{width_g}) for the remaining scenarios. The radiative decay in
scenarios 3 and 5 have considered $f_0$ resonance as quarkonium and tetraquark including Vector Meson Dominance as discussed
in Ref. \cite{Jacosa}.

\begin{figure}[ht]
\includegraphics[scale=0.25]{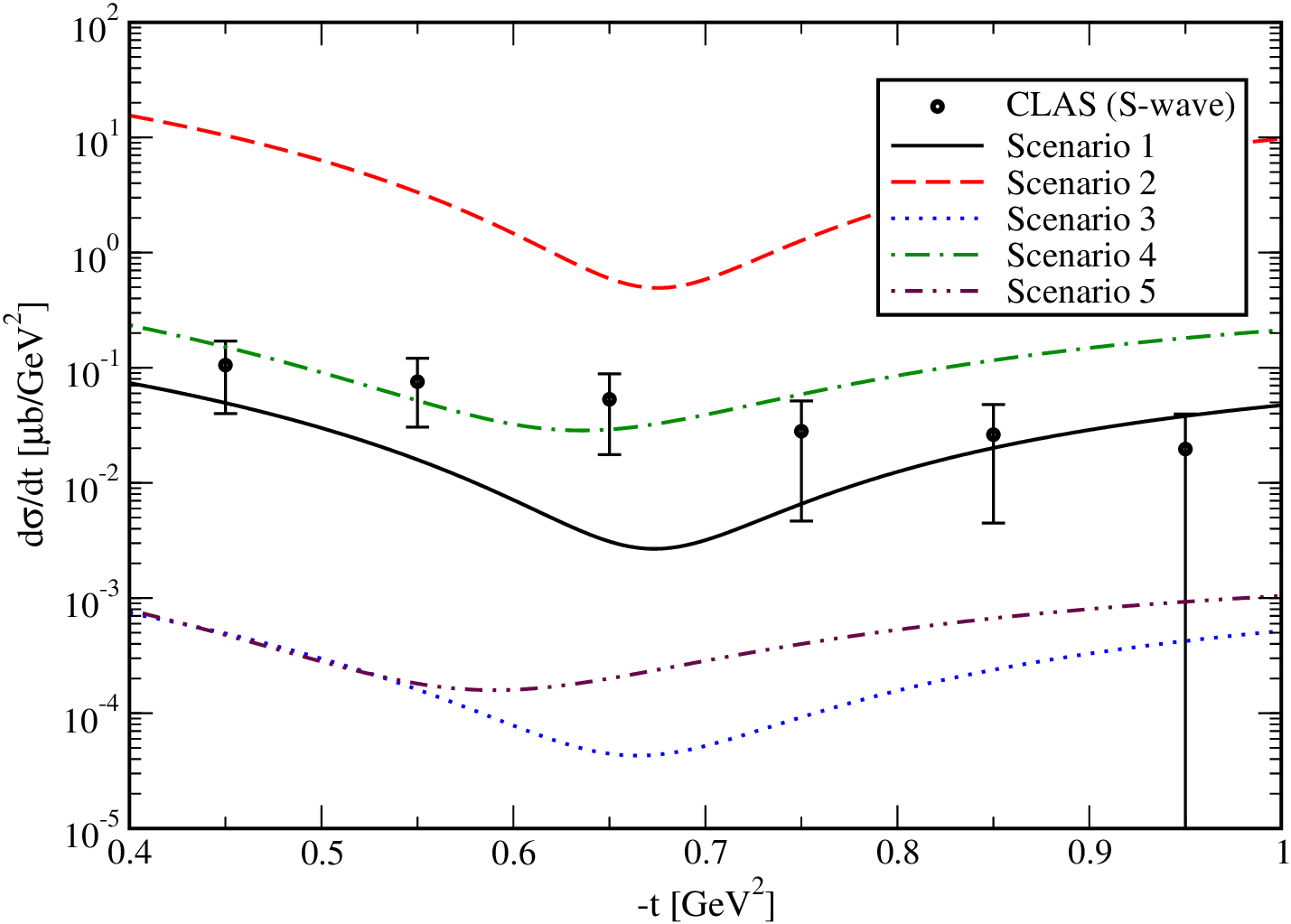}
\includegraphics[scale=0.25]{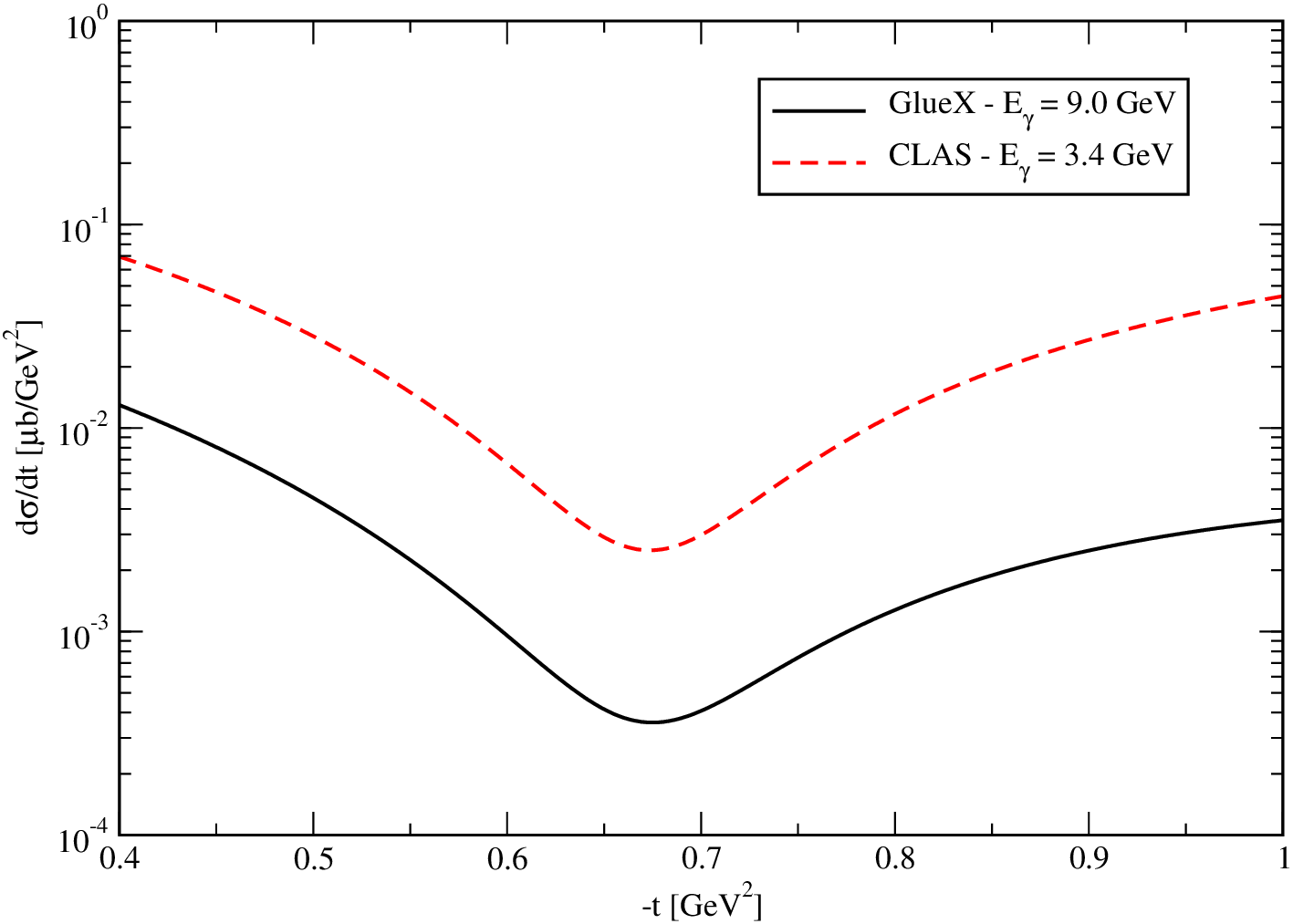}
\caption{The $S$-wave differential photoproduction cross section  for $f_0(980)$ photoproduction as a function of
momentum transfer squared at CLAS experiment energy $E_{\gamma}=  3.4$ GeV (left panel). We show a comparison between
the differential cross section for $f_0(980)$ in scenario 1 at CLAS and GlueX energies (right panel).}
\label{fig:1}
\end{figure}

The partial S-wave differential cross sections for the $f_0(980)$ are presented in Fig. \ref{fig:1} left panel at $E_{\gamma} = 3.4\pm 0.4$ GeV
and integrated in the $M_{\pi\pi}$ mass range $0.98\pm 0.04$ GeV. The dip at $ -t \approx 0.7$ GeV$^2$ related to the reggeized meson
($\rho$ and $\omega$) exchange. The scenarios 1 and 4 are represented by the solid and dot-dashed lines, respectively. Both of them
fairly reproduces the trend of CLAS data. The remaining scenarios are above or below the CLAS data points by a factor of 50.

The theoretical predictions are compared to the CLAS analysis at Jafferson Lab \cite{CLASf0}, where the $\pi^+\pi^-$ photoproduction
at photon energies between 3.0 and 3.8 GeV has been measured in the interval of momentum transfer squared $0.4 \leq |t|\leq 1.0$
GeV$^2$. There, the first analysis of S-wave photoproduction of pion pairs in the region of the $f_0(980)$ was performed. The
interference between $P$ and $S$ waves at $M_{\pi\pi}\approx 1$ GeV clearly indicated the presence of the $f_0$ resonance. The exclusive
reaction $\gamma p \rightarrow pf_0$ was measured through the most  sizable decay mode which is $f_0(980)\rightarrow \pi^+\pi^-$. 

In Figure \ref{fig:1} right panel we show a comparison between the $S$-wave differential photoproduction cross section  for $f_0(980)$ in the CLAS
($E_\gamma = 3.4$ GeV) and GlueX ($E_\gamma = 9$ GeV) energies. In this plot we consider the coupling $g_{K\bar{K}} = 0.4$, 
$g_{\pi\pi} = 1.31$ GeV and the $g_S$ coupling for senario 1. The results in Fig. \ref{fig:1} right panel shown that the differential cross section
for $E_{\gamma}=  9$ GeV is about an order of magnitude smaller than for $E_{\gamma}=  3.4 $ GeV.

\begin{figure}[ht]
\includegraphics[scale=0.24]{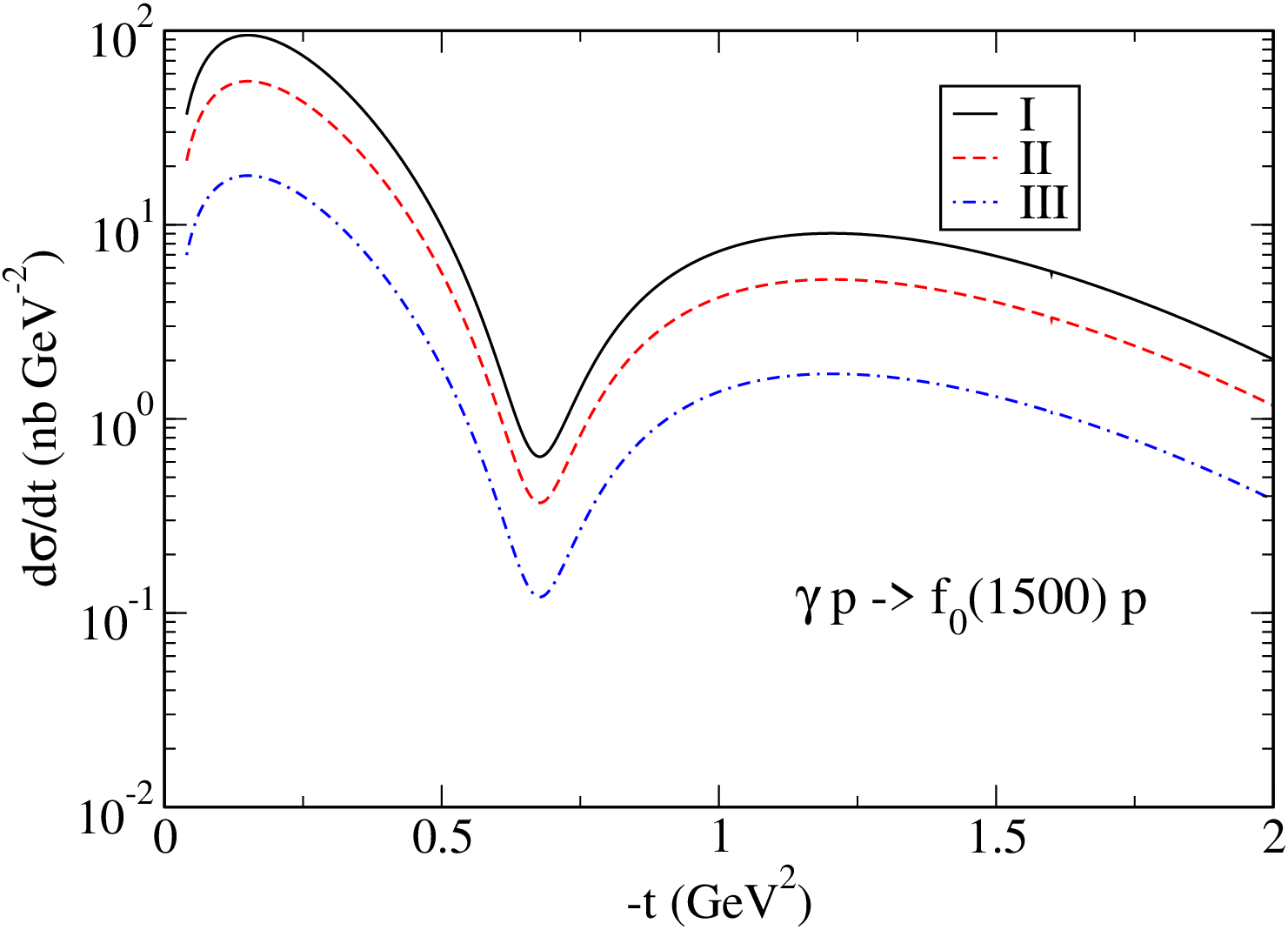}
\includegraphics[scale=0.24]{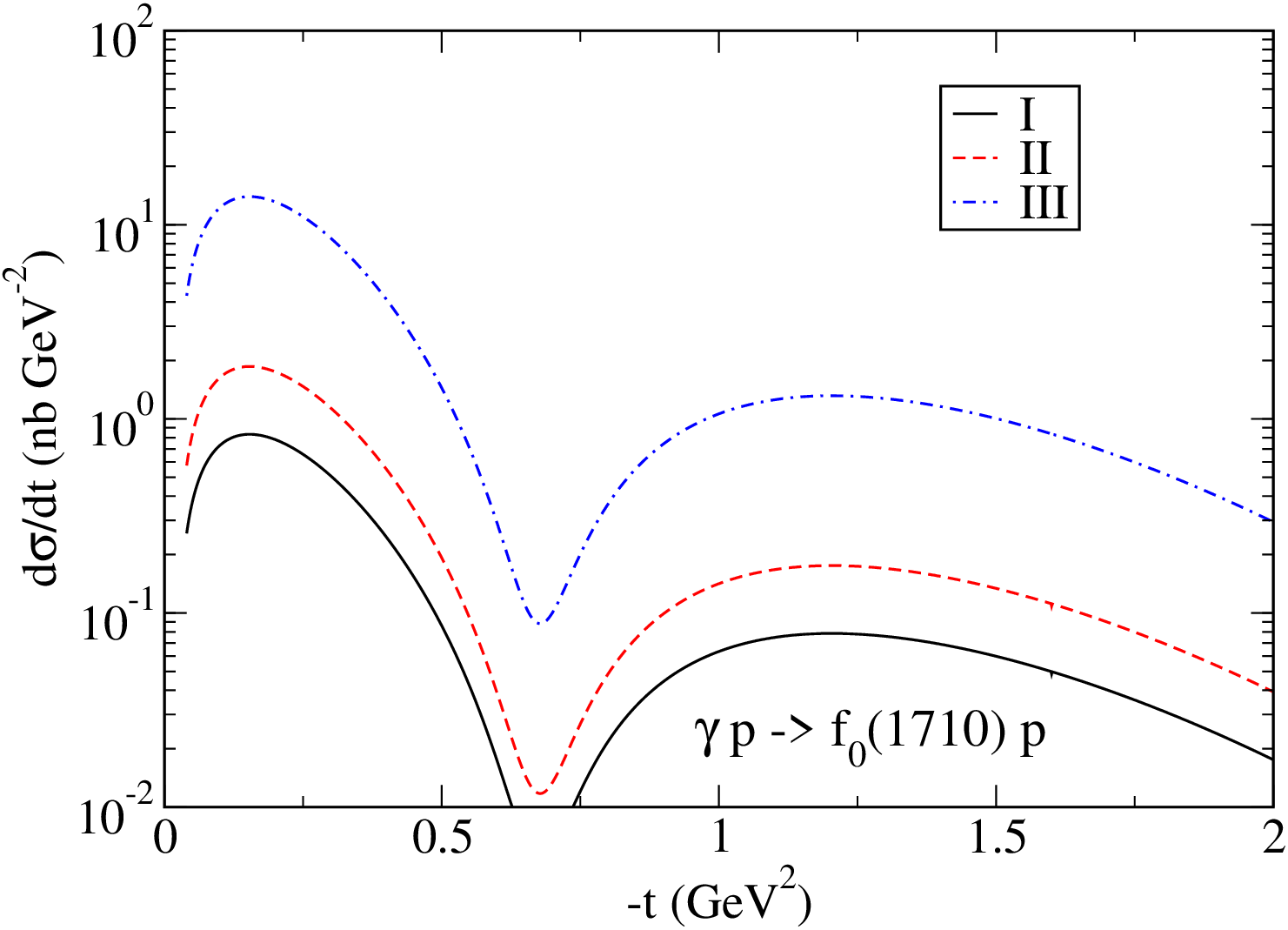}
\caption{Differential photoproduction cross section on proton for $f_0(1500)$ (left panel) and $f_0(1710)$ (right panel)
at GlueX energy $E_{\gamma}=9$ GeV.
The results for the distinct  three mixing scenarios are presented: I (solid line), II (dashed line) and III
(dot-dashed line).}
\label{fig:2}
\end{figure}

The differential cross sections for $f_0 (1500)$ and $f_0(1710)$ are presented in Fig. \ref{fig:2} at $E_{\gamma} = 9$
GeV, and showing the consequences of distinct mixing scenarios. In the
scenario (I) the cross section is higher the other scenarios. That is, a light glueball mass implies a larger cross
section for the $f_0 (1500)$ mesons. On the other hand, the inverse situation occurs for the $f_0(1710)$ mesons where
the large cross section comes from the heavy glueball mass component. The cross sections
reflect directly the radiative decay widths as can be verified from simple inspection of Table 1 of Ref. \cite{SM}. 
For completeness, the integrated cross sections for photoproduction of the scalars on protons at $E_\gamma$ = 9 GeV
are given in Table \ref{tab2} for light (I), medium (II) and heavy (III) glueball masses.

\begin{table}[t]
\begin{center}
\begin{tabular} {|l|c|c|c|}
\hline
Scenario & (I)  & (II)  & (III)  \\
\hline
\hline
 $f_0(1500)$ &  34.98 &  20.25 &  6.61 \\
\hline
$f_0(1710)$ &  0.30 & 0.68 &  5.08 \\
\hline
\end{tabular}
\end{center}
\caption{\it Integrated photoproduction cross sections in nanobarns on protons
at $E_\gamma = 9$ GeV for the three different mixing scenarios:
light (I), medium-weight (II) and heavy glueball (III) (see text).}
\label{tab2}
\end{table}

In summary, we have studied the photoproduction of $f_0(980)$ resonance for photon energies considered in the CLAS experiment, 
$E_{\gamma} = 3.4\pm 0.4$ GeV and in the GlueX experiment, $E_{\gamma}=  9$ GeV. It provides a test for current understanding of the
nature of the scalar resonances. We have calculated the differential cross sections as function of effective masses and momentum transfers.
The effect of distinct scenarios in the calculation of coupling  $S\rightarrow V\gamma$ were investigated. This investigation shows that
we need to known more precisely the radiative decay rates for $f_0(980)\to \gamma V$ which are important to this calculation. Our
predictions of the cross sections are somewhat consistent with the experimental analysis from CLAS Collaboration. The present experimental
data are able to exclude some possibilities for the  $S\rightarrow V\gamma$  coupling. 
An estimation of the differential cross section for the GlueX experiment is also presented.
We also have studied the photoproduction of the $f_0(1500)$ and $f_0(1710)$ resonances for
photon energies relevant for the GlueX experiment at photon energy of 9 GeV. It would provide novel tests for our
understanding of the nature of the scalar resonances and about current ideas on glueball and $q\bar{q}$ mixing.
The meson differential and integrated cross sections were evaluated and the effect of distinct mixing scenarios
were investigated. Although large backgrounds are expected, the signals could be visible by considering only the
all-neutral channels, that is their decays on $\pi^0\pi^0$, $\eta^0\eta^0$ and $4\pi^0$. The theoretical
uncertainties are still large, with $f_0(1500)$ the more optimistic case. Finally, an experiment in nuclei would
also lead to the $f_0$ and $a_0$ excitation mostly from the collision of protons with protons.

\section*{Acknowledgments}
This research was supported by CNPq and FAPERGS, Brazil.


\end{document}